\documentclass{ws-ijmpe}

\begin{document}

\markboth{D. Dutta and K. Hafidi}{The Search for the Onset of Color Transparency: A Status Report}

\catchline{}{}{}{}{}

\title{THE SEARCH FOR THE ONSET OF COLOR TRANSPARENCY: A STATUS REPORT}

\author{\footnotesize DIPANGKAR DUTTA}

\address{Department of Physics and Astronomy, Mississippi State University, Mississippi State, Mississippi 39762, USA \\
d.dutta@msstate.edu}

\author{KAWTAR HAFIDI}

\address{Physics Division, Argonne National Laboratory, Argonne, Illinois 60439, USA \\
kawtar@anl.gov}

\maketitle

\begin{history}
\received{(received date)}
\revised{(revised date)}
\end{history}

\begin{abstract} 
Color Transparency refers to the vanishing of the hadron-nucleon interaction for hadrons produced inside a nucleus in high momentum exclusive processes. We briefly review the concept behind this unique Quantum Chromo Dynamics' phenomenon, the experimental search for its onset and the recent progress made at intermediate energies.
\end{abstract}

\section{Introduction}
The concept of Color Transparency (CT) was introduced almost three decades ago by Mueller and Brodsky \cite{Muller82}, as a distinctive property of QCD, related to the presence of color degrees of freedom underlying strongly interacting matter. It refers to the suppression of the interaction between small size color singlet wave packets and hadrons, due to the cancellation of the color fields of Quantum Chromo Dynamics (QCD)~\cite{low75}. The basic idea is that at sufficiently high momentum transfer, the quarks inside a hadron, each of which would normally interact very strongly with nuclear matter, could form an object of reduced transverse size. This small size object should be `color neutral' outside of its small radius in order not to radiate gluons. And if this compact size is maintained for distances comparable to the size of the nucleus it would  pass through the nuclear medium without further interactions. Hence, nuclear transparency, which is defined as the ratio of the cross section per nucleon for an exclusive scattering process on a bound nucleon in the nucleus to the cross section for the same process on a free nucleon, is the observable of choice in searching for the onset of CT phenomenon. A combination of the effects described above would lead to an increase in the measured nuclear transparency with energy. A clear signature for the onset of CT would involve a rise in the nuclear transparency as a function of momentum transfer. The phenomenon of CT is essential for ensuring the Bjorken scaling in deep inelastic scattering at high energies. A similar phenomenon occurs in QED, where an $e^+e^-$ pair of small transverse size has a small cross section determined by its electric dipole moment \cite{perkins}. In QCD, a $q \bar{q}$ or $qqq$ system can act as an analogous small color dipole moment.

Although, CT was first discussed in the context of perturbative QCD, however, later works \cite{CTreview} have indicated that this phenomenon also occurs in a wide variety of models which feature non-perturbative reaction mechanisms. Recently, CT has also been discussed in the context of QCD factorization theorems. These factorization theorems were, over the last few years, derived for various deep inelastic exclusive processes \cite{factor1,factor2,factor3,factor4}, and are intrinsically related to accessing the Generalized Parton Distributions (GPDs), introduced by Ji and Radyushkin \cite{gpd1,gpd2}. The discovery of these GPDs and their connection to  exclusive cross sections has made it possible in principle to rigorously map out the complete nucleon wave functions themselves. Presently, experimental access to such GPDs is amongst the highest priorities in hadronic physics. 

For exclusive processes such as meson electroproduction on a baryon, upon absorbing the virtual photon the meson and the baryon move fast in opposite directions. It has been suggested~\cite{strikman} that the outgoing meson maintains a small transverse size, which results in a suppression of soft interactions (multiple gluon exchange) between the meson-baryon systems moving fast in opposite directions and thereby leading to factorization. Consequently, factorization is rigorously not possible without the onset of the CT phenomenon \cite{strikman}. The underlying
 assumption here is that in exclusive ``quasi-elastic'' hadron production, the hadron is produced at small inter-quark distances. However, just the onset of CT is not enough, because higher-twist contributions such as quark transverse momentum contributions can be large at lower four momentum transfer squared ($Q^2$) which could lead to breakdown of factorization~\cite{vgg,hermes}. Therefore, the onset of CT in hadron production is a necessary but not sufficient condition for the validity of factorization. It should be noted that it is still uncertain at which ($Q^2$) value one will reach the factorization regime. An unambiguous observation of CT would be the first step in determining the onset of the factorization regime.    


Finally, it has been predicted \cite{ralston} that for exclusive processes in a nuclear medium, 
large quark separations will tend not to propagate significantly in the strongly interacting medium. Configurations of small quark separations, on the other hand, will propagate with small attenuation. This phenomenon is termed nuclear filtering, and is the complement of CT phenomenon. When such nuclear filtering occurs, the nuclear medium should eliminate the long distance amplitudes. If this prediction holds, in the large $A$ limit, exclusive reactions in the nuclear medium would be perturbatively calculable at lower energy scales compared to the corresponding free nucleon case.

\section{CT at High Energies}
At high energies the CT phenomenon arises from the fact that, exclusive processes on a nucleus at high momentum transfer preferentially select the color singlet small transverse size configuration, which then moves with high momentum in a ``frozen mode'' through the nucleus. The interaction between the small transverse size configuration and the nucleon is strongly suppressed because the gluon emission amplitudes arising from different quarks cancel. This suppression of the interactions is one of the essential ingredients needed to account for Bjorken scaling in deep-inelastic scattering at small $x$~\cite{Frankfurt88}. Thus the discovery of Bjorken scaling in deep-inelastic scattering 
can be considered as the first indirect evidence for CT at high energies.

The first direct evidence for CT at high energies came from the $A$ dependence of $J/\psi$ production by real photons in the energy range of 80 - 190 GeV. The measurements were performed at Fermilab \cite{Sokoloff:1986prl}
using H, Be, Fe, and Pb targets. These processes select the small transverse size configurations in the $J/\psi$ wave function by employing the decrease of the transverse separation between $q$ and $\bar q$ forming the $J/\psi$. The measured cross section can be parametrized as $\sigma_A = \sigma_1A^{\alpha}$, where $\sigma_1$ is a constant independent of the nucleus mass number $A$. From a vector-meson dominance model one expects~\cite{Bauer:1977iq} $\alpha = 4/3$ and the experiment measured $\alpha = 1.4 \pm 0.06 \pm 0.04$ for the coherently produced $J/\psi$. This result can be interpreted as due to CT at high energies.

\begin{figure}[th]
\centerline{\psfig{file=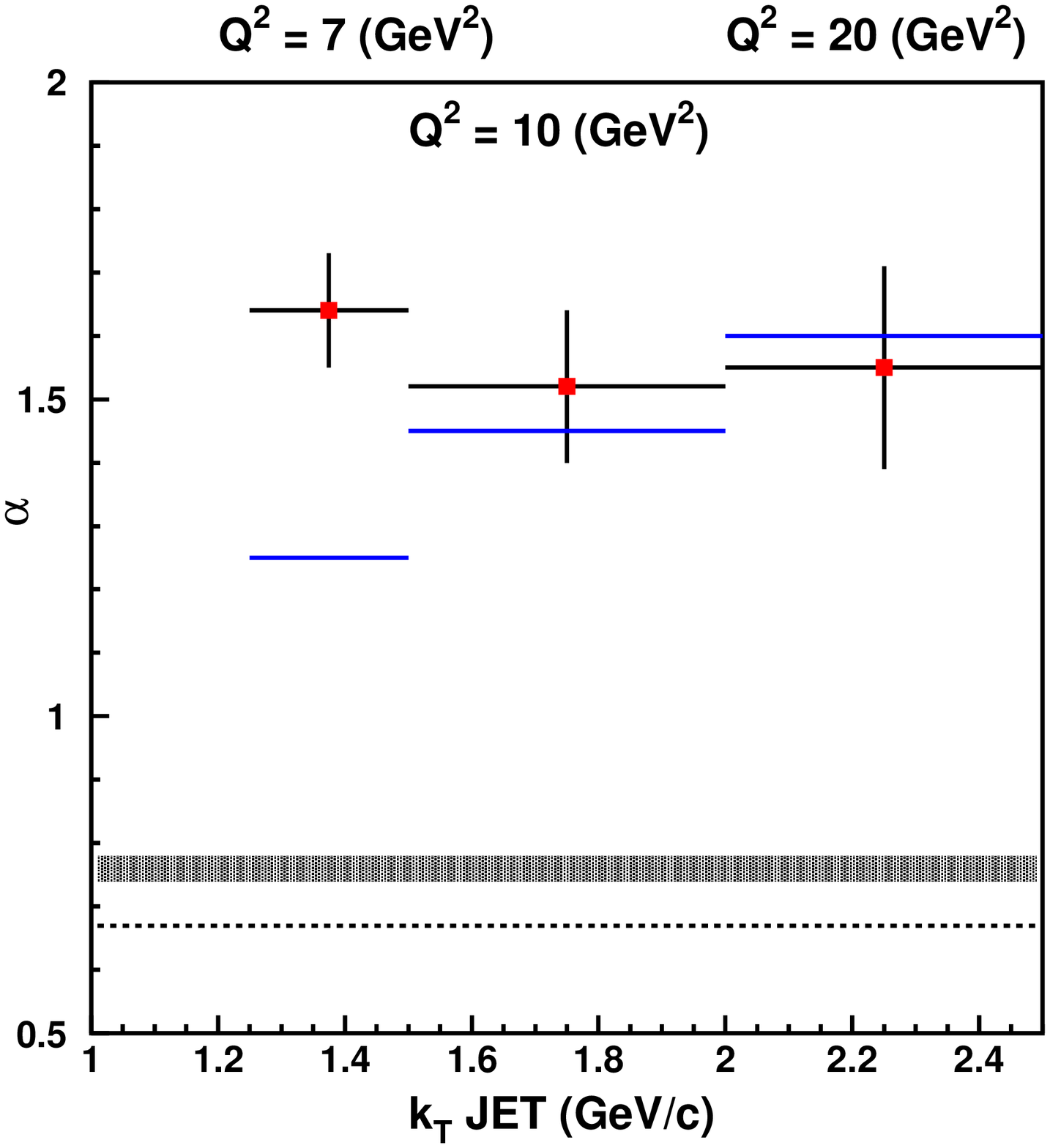,width=8cm}}
\vspace*{8pt}
\caption{The results from Fermilab experiment E791 showing the values of $\alpha$ obtained from parametrization of the E791 dijet cross section as $\sigma = \sigma_{0} A^{\alpha}$. The data are shown as red points along with the quadrature sum of statistical and systematic errors and the $k_{T}$ bin size. The blue lines are the CT predictions~\protect\cite{Frankfurt:1993it} and the dark band is $\alpha$ observed in pion-nucleus interactions, and the dashed line is the naive expectation of $\alpha = 2/3$.} 
\label{fig:dijet3}
\end{figure}

CT at high energies was also observed in FNAL experiment E791~\cite{Aitala:2000hc}, which measured the diffractive dissociation of 500 GeV/c pions into di-jets when coherently scattering from carbon and platinum targets. The per-nucleon cross section for di-jet production is parametrized as $\sigma = \sigma_{0} A^{\alpha}$, and the values of $\alpha$ obtained from the experiment E791 are shown in Figure~\ref{fig:dijet3} along with the CT 
predictions of \cite{Frankfurt:1993it}. These results confirm the predicted strong increase of the cross section with A:  $\sigma \propto A^{1.61\pm 0.08}$ as compared to the predicted $\sigma \propto A^{1.54}$, and the dependence of the cross section on the transverse momentum of each jet with respect to the beam axis ($k_t$) indicating the preferential selection of the small transverse size configurations in the projectile.

These experiments have unambiguously established  the presence of small size  $q\bar q$ configurations in light mesons and show that at  transverse  separations d $\le$  0.3 fm pQCD reasonably describes small dipole-like $q\bar q$ - nucleon interactions for $10^{-4} < x < 10^{-2}$. Thus, Color transparency is well established for the small dipole interaction with  nuclei for $x \sim 10^{-2}$. However, these high energy experiments do not provide
any information about the appropriate energy regime for the onset of CT.  

\section{Search for the Onset of CT at Intermediate Energies}
At intermediate energies, in addition to the preferential selection of the small size configuration, the expansion time of the interacting small size configuration often called pre-hadron is also very important. The largest longitudinal  distance for which coherence effects are still present (coherence lengths) is determined by the smallest characteristic internal excitation energies of the hadron. At intermediate energies these longitudinal distance scales are not large enough for the small size configuration to escape without interaction and this leads to strong suppression of the color transparency effect~\cite{FLFS88,jm90}. In this energy regime, the interplay between the selection of the small transverse size and its subsequent expansion determine the energy scale for the onset of CT. Estimates~\cite{FLFS88,jm90} show that for the case of the knock out of a nucleon, the coherence is completely lost at distances $\mbox{l}_c \sim 0.4\div 0.6 ~\mbox{fm}\cdot  \, \mbox{p}_h$, where $\mbox{p}_h$ is the momentum of the final hadron measured in GeV/c. Hence even if a nucleon is produced in a small size configuration it has to have momentum significantly larger than $p_N[GeV] > r_{NN}/0.5 \sim 4$ GeV for a significant change of the transparency (here $r_{NN} \sim 2$ fm is the typical mean free path of a nucleon in the nucleus). This corresponds to $Q^2 > 8$ (GeV/c)$^2$.

\subsection{Early experiments}
\subsubsection{Quasi-elastic Proton Scattering on Nuclei}
\begin{figure}[th]
\centerline{\psfig{file=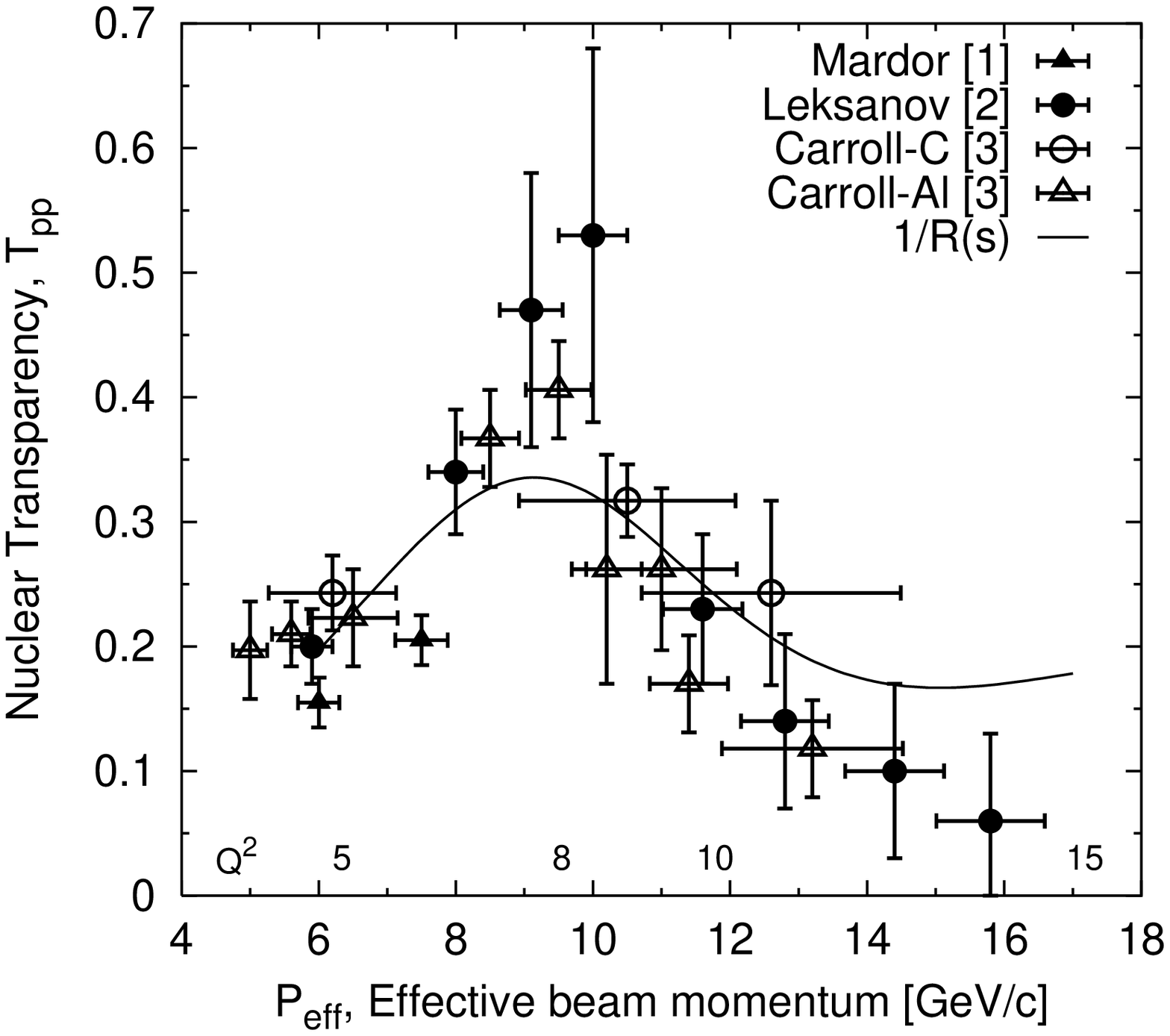,width=6.75cm}\psfig{file=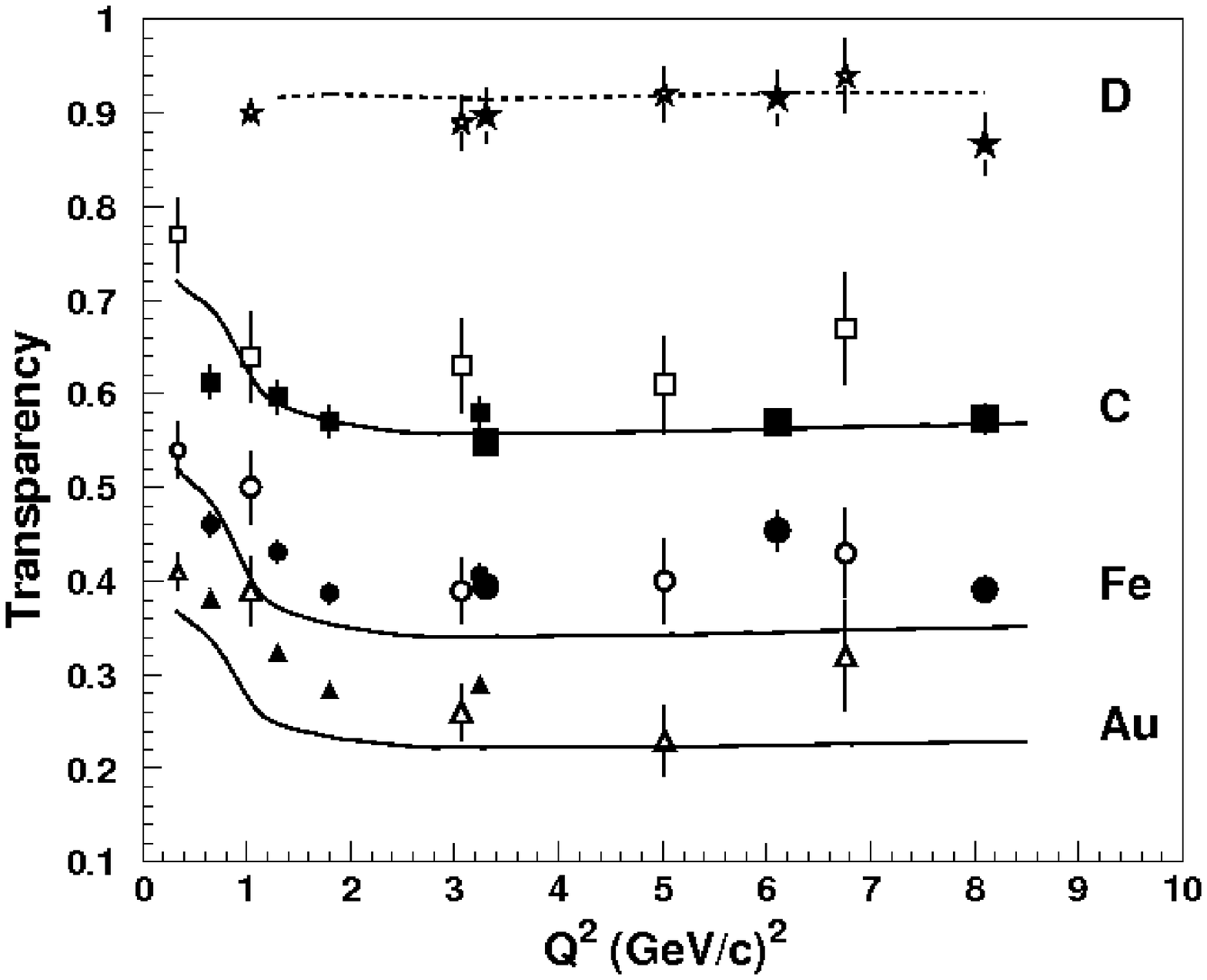,width=6.75cm}}
\vspace*{8pt}
\caption{(Left panel) The nuclear transparency values for $^{12}$C and $^{27}$Al (scaled by $(27/12)^{1/3})$ 
versus the effective beam momentum. The solid curve is the inverse of $R(s)$ as defined in the text~\protect\cite{Aclander:2004zm}. (Right panel) A compilation of transparency for $(e,e'p)$ quasi-elastic scattering~\protect\cite{Garrow:2002} from D (stars), C (squares), Fe (circles), and Au (triangles). Data from the two JLab experiments~\protect\cite{Abbott:1998,Garrow:2002} are shown as solid points. The previous SLAC data~\protect\cite{Makins:1994} are shown as large open symbols, and the previous Bates data~\protect\cite{Garino:1992} are shown as small open symbols, at the lowest $Q^2$ on C, Ni, and Ta targets, respectively. The errors shown for the JLab measurements (solid points) include statistical and the point-to-point systematic (2.3\%) uncertainties, but do not include model dependent systematic uncertainties or normalization-type errors. The error bars for the other data sets include their net systematic and statistical errors. The solid curves shown from $0.2 < Q^2 < 8.5 $(GeV/c)$^2$ are Glauber calculations from Ref.~\protect\cite{vijay:92}. In the case of D, the dashed curve is a Glauber calculation from M. Sargsian~\protect\cite{Misak:95}.} 
\label{fig:proton}
\end{figure}

The first attempt to measure the onset of CT at intermediate energies was undertaken at the Brookhaven National Lab (BNL) and it used large angle proton knockout $A(p,2p)$ reaction~\cite{Carroll:88}. In that experiment large 
angle $pp$ and quasi-elastic $(p,2p)$ scattering were simultaneously measured in hydrogen and 
several nuclear targets, at incident proton momenta p$_p$ of 6 - 12 GeV/c. The nuclear transparency
 was extracted from the ratio of quasi-elastic cross section from a nuclear target to the free $pp$ elastic cross section. The transparency was found to increase as predicted by CT, between 6 - 9.5 GeV/c but decreased between 9.5 and 12 GeV/c. A dedicated followup experiment EVA~\cite{Leksanov:01} extended these measurements to 14.4 GeV/c. The final results from both experiments~\cite{Aclander:2004zm} are shown in Fig~\ref{fig:proton} (left). In addition to the energy dependence of the transparency, the angular dependence (80 $< \theta_{c.m.} <$ 90) of the 
transparency was also measured.

The initial rise in transparency between $\mbox{p}_p$= 6 - 9.5 GeV/c is consistent with the 
selection of a small size configuration and its subsequent expansion over distances comparable 
to the nuclear radius. However, the non-monotonous energy dependence of the transparency, is 
a problem for all current models with just CT. Two possible explanations have been suggested 
for the observed energy dependence above $\mbox{p}_p$= 9 GeV/c. One ascribes the energy 
dependence to an interference between a hard amplitude, which dominates the high energy 
$pp$ elastic scattering cross section, and a soft amplitude arising from higher order radiative 
process, also known as Landshoff mechanism~\cite{Ralston:1988rb,Jain:1995dd}. The $pp$ elastic 
scattering cross section near $90^{\circ}_{c.m.}$ degrees varies with 
c.m.energy $(s)$ as;
\begin{equation}
\frac{d\sigma}{dt_{pp}}(\theta = 90^{\circ}_{c.m.}) = R(s) s^{-10}
\end{equation}
In the Landshoff mechanism picture, the long-ranged part of the amplitude is attenuated 
by the nuclear matter, and the interference disappears for the nuclear cross section and 
hence the energy dependence of the transparency should be the inverse of $R(s)$, as shown by the solid curve in Fig.~\ref{fig:proton} (left).
The second explanation~\cite{Brodsky:1987xw} suggests that the energy dependence of the $pp$ 
elastic scattering cross section scaled by $s^{-10}$ corresponds to a resonance or threshold 
for a new scale of physics, such as charmed quark resonance or other exotic QCD multi-quark 
states.   

\subsubsection{Quasi-elastic Electron Scattering on Nuclei}
Compared to hadronic probes, the weaker electromagnetic probe samples the complete nuclear volume. Moreover, the fundamental electron-proton scattering cross section is smoothly varying and is accurately known over a
wide kinematic range and detailed knowledge of the nucleon energy and momentum distribution inside a variety of nuclei have been measured extensively in low energy experiments. The advantages of electronuclear reactions was immediately recognized following the BNL $(p,2p)$ experiments and an effort to measure CT using electron scattering was launched.

In quasi-elastic $(e, e'p)$ scattering from nuclei the electron scatters from a single proton which is moving due to its Fermi momentum~\cite{Frulani:1984}. In the plane wave impulse approximation (PWIA) the proton is ejected without final state interactions with the residual (A-1) nucleons. The measured $A(e, e'p)$ cross section would be 
reduced compared to the PWIA prediction in the presence of final state interactions, where  the proton can scatter both elastically and inelastically from the surrounding nucleons as it exits the nucleus. The deviations from the simple PWIA expectation is used as a measure of the nuclear transparency. In the limit of complete color transparency, the final state interactions would vanish and the nuclear transparency, defined as the ratio of the measured to the PWIA 
cross section, would approach unity. 
 
In the conventional nuclear physics picture, one expects the nuclear transparency to show the same energy dependence as the energy dependence of the Nucleon-Nucleon {\it (NN)} cross section. Other effects such as short-range correlations and the density dependence of the {\it NN} cross section will affect the absolute magnitude of the nuclear transparency but have little influence on the $Q^2$ dependence of the transparency. Thus the onset of CT would manifest as a rise in the nuclear transparency as a function of increasing $Q^2$. However, even a conclusive experimental observation of a rise in nuclear transparency with increasing $Q^2$, may not necessarily be an 
unambiguous observation of CT~\cite{kopel:96}, because such a rise can also be caused by the diffractive production of inelastic intermediate states by the knocked-out proton while it propagates through the medium since in that case both the virtual photon energy $\nu$ and $Q^2$ are correlated.

The $(e,e'p)$ reaction is expected to be simpler to understand than the $(p,pp)$ reaction and is 
not affected by either of the two explanations proposed to account for the observed energy dependence of nuclear transparency in $(p,pp)$ reactions discussed earlier. The first electron scattering experiment to look for the onset of CT was the NE-18 $A(e,e'p)$ experiment at SLAC~\cite{Makins:1994}. This experiment yielded distributions in missing energy and momentum completely consistent with conventional nuclear physics and the extracted 
transparencies exclude sizable CT effects up to $Q^{2}$ = 6.8 (GeV/c)$^2$ in contrast to the results from the $A(p,2p)$ experiments~\cite{Carroll:88}. Later experiments with greatly improved statistics and systematic uncertainties compared to the NE-18 experiment~\cite{Makins:1994}, and with increased $Q^2$ range was carried out at 
JLab~\cite{Abbott:1998,Garrow:2002}.

A compilation of the measured transparency $T(Q^2)$ values from all electron scattering experiments is presented in Fig.~\ref{fig:proton} (right). The results show no $Q^2$ dependence in the nuclear transparency data
above $Q^2$ of 2 (GeV/c)$^2$. The energy dependence below $Q^2$ = 2 (GeV/c)$^2$ is consistent with the energy dependence of the $p$-nucleon cross section. Above $Q^2$ = 2 (GeV/c)$^2$,  excellent constant-value fits were obtained for the various transparency results. In Fig.~\ref{fig:proton} (right) the measured transparency is compared with the results from correlated Glauber calculations, including rescattering through third order~\cite{vijay:92} (solid curves for $0.2 < Q^2 < 8.5$ (GeV/c)$^2$). In the case of deuterium the dashed curve shows a generalized Eikonal approximation calculation~\cite{FGMSS:95}, which coincides with a Glauber calculation for small missing momenta~\cite{Misak:95}. Although these models can describe the $Q^2$ dependence of the nuclear transparencies, the absolute magnitude of the transparencies are under-predicted for the heavier nuclei. This behavior persists even after accounting for the model-dependent systematic uncertainties. The $Q^2$ independence of the transparencies may also result from canceling effects in the hard electron-proton scattering and CT. 

In addition to the  $Q^2$ dependence, the nuclear mass number $A$ dependence of the nuclear transparency was also studied by parametrization of the transparency to the form $T=c~A^{\alpha(Q^2)}$. Within uncertainties, the constant $c$ is found to be consistent with unity as expected and the parameter $\alpha (Q^2)$ to exhibit no $Q^2$ dependence up to $Q^2$ = 8.1 (GeV/c)$^2$ with a nearly constant value of $\alpha$ = - 0.24 for $Q^2 > $ 2.0 (GeV/c)$^2$. This is also consistent with conventional nuclear physics calculations using the Glauber approximation.

The existing world data rule out any onset of CT effects larger than 7\% over the $Q^2$ range from 2.0 to 8.1 (GeV/c)$^2$, with more than 90\% confidence level. The $(e,e'p)$ data seem to suggest that a $Q^2$ of 8 (GeV/c)$^2$ is not large enough to overcome the expansion of the small transverse size objects selected in the hard $e p$ scattering process.

\subsubsection{Meson Production Experiments}
It is expected that it is more probable to reach the CT regime at lower energies for the interaction/production  of mesons than for baryons since only two quarks have to come close together and a quark-antiquark pair is more likely to form a small size  object~\cite{Blaettel:1993rd}. Further, it is important to note that the unambiguous observation of the onset of CT is a critical precondition for the validity  of the factorization theorem for meson production~\cite{Collins:1996fb}. This  is because in the regime where CT applies, the outgoing meson retains a small transverse size (inter-quark distance) while soft interactions like multiple gluon exchange between the meson produced from the hard interaction and the baryon are highly suppressed. QCD factorization in hard exclusive processes is thus rigorously not possible without the onset of CT~\cite{gpdct}.

As described earlier, the $J/\psi $ coherent and quasi-elastic photoproduction experiments did find a weak absorption of $J/\psi$ indicating presence of CT. Evidence for CT was also observed in the coherent diffractive dissociation of 500~GeV/c negatively charged pions into di-jets. There were also hints for CT in  several $\rho$-meson production 
experiments~\cite{Adams:1995,Airape:2003}. However all of these high energy experiments did not 
have good enough resolution in the missing mass to suppress hadron production at the nucleus 
vertex, making interpretation of these experiments somewhat ambiguous. Moreover, these high 
energy experiments do not tell us anything about the onset of CT and therefore one cannot learn about the interaction and evolution of these special objects in the nuclear medium.

\subsection{Recent Experiments at Jefferson Lab}
A recent high resolution experiment of pion production reported evidence for the onset 
of CT~\cite{Clasie:2007} in the process $eA\to e\pi^+ A^*$. New results for the $\rho$-meson production at JLab have also confirm the early onset of CT in mesons~\cite{ElFassi:2012nr}. The pion electroproduction and rho experiments together conclusively demonstrate the onset of CT in the few GeV energy range. These experiments are discussed below.

\subsubsection{Pion Production Experiments}
\label{sec:pion}

{\it (i) Pion Photoproduction}\\
The onset of CT was first explored in a pion photoproduction experiment at JLab. In this experiment, nuclear transparency of the $\gamma n \rightarrow \pi^{-} p$ process was measured as a ratio of pion photoproduction cross section from $^4$He to $^2$H~\cite{Dutta:2003}. The elementary hadron-nucleon cross-sections along with the exact nuclear ground state wave function for $^4$He~\cite{Arriga:96} can be used to carry out precise calculations of the nuclear transparency~\cite{Gao:96}. Therefore, measurements of nuclear transparency from $^4$He nuclei  constitute a benchmark test of traditional nuclear calculations. In addition, light nuclei such as $^4$He are predicted to be better for the onset of CT phenomenon because of their relatively small nuclear sizes, which are smaller than the expansion length scales of the small size object~\cite{FLFS88}.


The photopion results on $^4$He appears to deviate from the traditional nuclear physics calculations at the higher energies. The slopes of the measured transparency obtained from the three points which are above the resonance region (above $E_{\gamma}$ = 2.25 GeV ) are in good agreement, within experimental uncertainties, with the slopes predicted by the calculations including CT~\cite{FLFS88} and they seem to deviate from the slopes predicted by the Glauber calculations~\cite{Gao:96} without CT at the $ 1 \sigma (2 \sigma)$ level for $\theta_{CM}^{\pi} = 70^{\circ} (90^{\circ})$. These data suggest the onset of behavior predicted for CT, but future experiments 
with significantly improved statistical and systematic precision are essential to confirm such conclusions.

~\\
\noindent
{\it (ii) Pion Electroproduction}\\
The first extensive study of the pion electroproduction on a number of nuclear targets 
($^1$H, $^2$H, $^{12}$C, $^{27}$Al, $^{63}$Cu and $^{197}$Au) was carried out at JLab in 2004. This experiment 
(piCT) made it possible for the first time to determine simultaneously the $A$ and $Q^2$ dependence of the pion
differential cross section for $Q^2$ = 1 - 5 (GeV/c)$^2$~\cite{Clasie:2007,Qian:2010}. The fraction of pions 
escaping from the nucleus is defined as the pion nuclear transparency.  In the quasi-free picture, the ratio of 
the longitudinal to transverse cross section from a bound proton inside the nucleus is expected to be the 
same as that from a free proton, this also provides the means to test the appropriateness  of the quasi-free 
approximation. Assuming the dominance of the quasi-free process, one can extract the nuclear 
transparency of the pions, by taking the ratio of the acceptance corrected cross sections from the 
nuclear target to those from the proton and/or deuteron. 

The piCT experiment verified the dominance of the quasi-free process by comparing the ratios of 
the longitudinal to transverse cross sections from nuclear targets with those obtained from a 
nucleon target. Within experimental uncertainties, the $\sigma_L/\sigma_T$ ratios were found to be 
independent of $A$~\cite{Qian:2010}. This can be viewed as a confirmation of the quasi-free 
reaction mechanism. Additionally, the restriction of $-t \le $ 0.5 GeV$^2$ minimized 
contributions from rescattering or multi-nucleon effects. The coherence length defined as the distance between the point where $\gamma^*$ converted to a $q\bar q$ and where $q\bar q$ interacts with a nucleon $|l_{in}|=(Q^2+M^2_{q\bar q}/2q_0)$ is small for the kinematics of the piCT experiment and varies weakly with $Q^2$, where $M_{q\bar q}$ is the mass of the $q\bar q$ pair and $q_0$ is the virtual photon energy. This simplifies  interpretation of the $Q^2$ dependence of the transparency as compared to the case of small x where  $l_{in}$ becomes comparable to the nucleus size.

\begin{figure}[th]
\centerline{\psfig{file=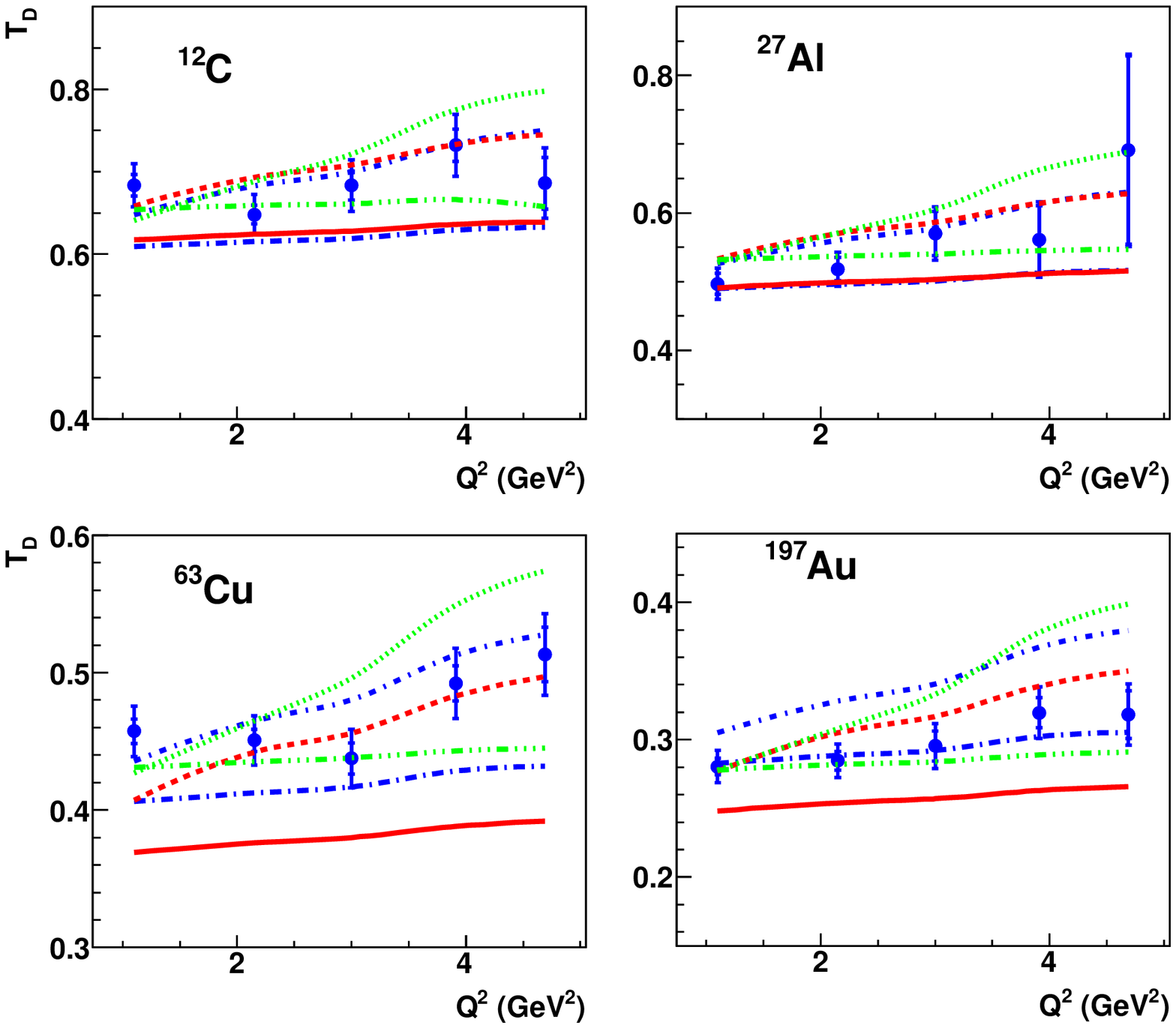,width=8.5cm}\psfig{file=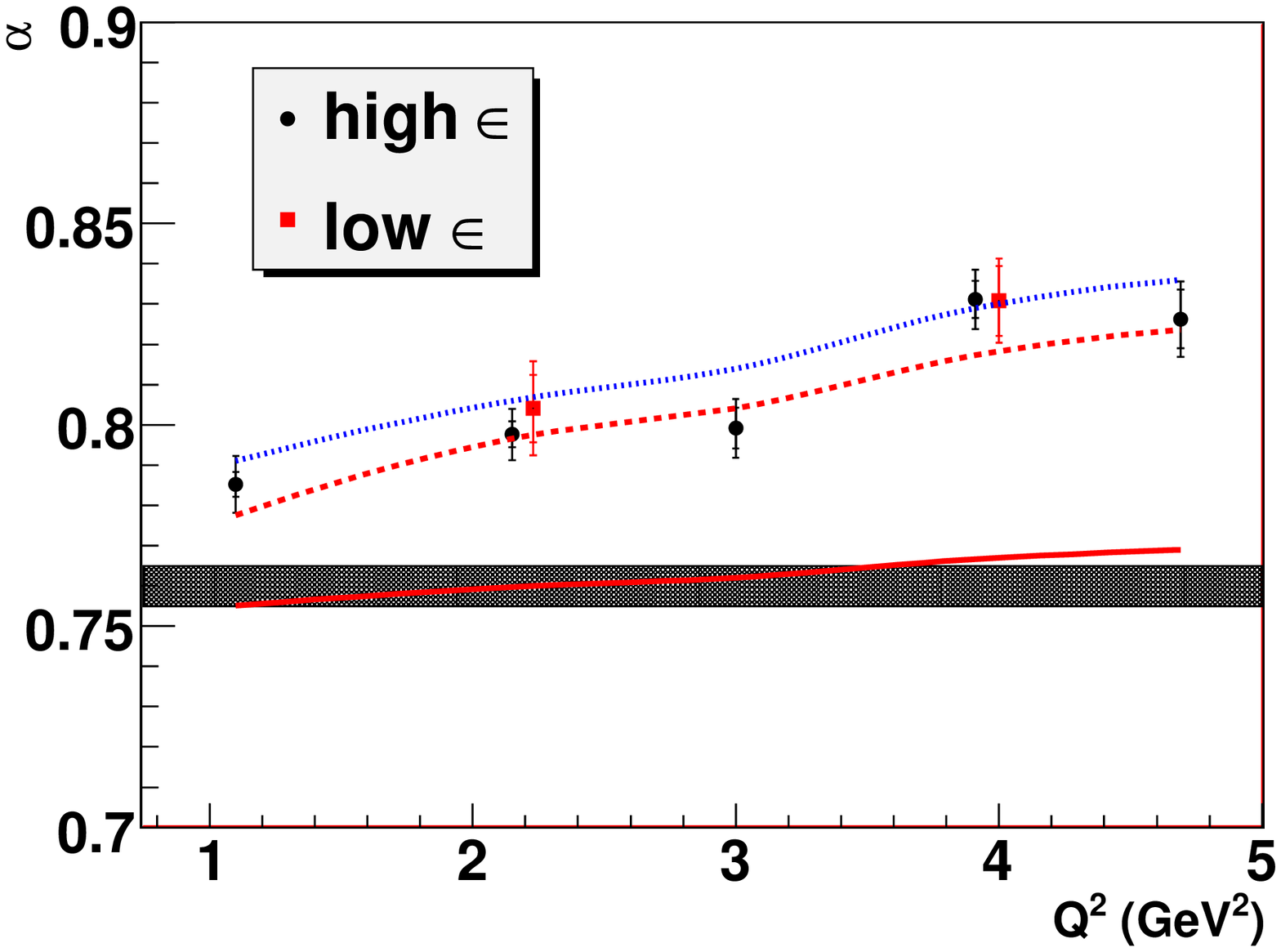,width=5cm}}
\vspace*{8pt}
\caption{(Left panel) Nuclear transparency vs $Q^2$ for $^{12}$C, $^{27}$Al, $^{63}$Cu and $^{197}$Au. The inner error bars are the statistical uncertainties and the outer error bars are the statistical and point-to-point systematic uncertainties added in quadrature. The solid circles (blue) are the high $\epsilon$ (virtual photon polarization) points, while the solid  squares (red) are the low $\epsilon$ points. The dashed and solid lines (red) are Glauber calculations  from Larson, {\it et al.}~\protect\cite{Larson:2006ge}, with and without CT, respectively. Similarly, the dot-short dash and dot-long dash lines (blue) are Glauber calculations with and without CT from Cosyn, {\it et al.}~\protect\cite{Cosyn:2006}. The dotted and dot-dot-dashed lines (green) are microscopic+ BUU transport calculations from Kaskulov {\it et al.}~\protect\cite{mosel2}, with and without CT, respectively. (Right panel) The parameter $\alpha (Q^2)$, as extracted from a fit of the nuclear transparency to the form $T=A^{(\alpha - 1)}$ (solid black circles). The inner error bars indicate the statistical uncertainties, and the outer error bars are the quadrature sum of statistical, systematic and modeling uncertainties. The hatched band is the value of $\alpha (Q^2)$ extracted from pion-nucleus scattering data~\protect\cite{caroll79}. The solid, dashed and dotted lines are $\alpha$ obtained from fitting the $A$-dependence of the theoretical calculations: the Glauber and  Glauber+CT calculations~\protect\cite{Larson:2006ge}, and the Glauber + CT (including short-range correlation effects) calculations~\protect\cite{Cosyn:2006}, respectively. The red circles show the $\alpha$ values extracted at the low $\epsilon$ kinematics.}
\label{fig:pict}
\end{figure}

The pion nuclear transparency was calculated as the ratio of pion electroproduction cross sections from the nuclear target to those from the proton~\cite{Clasie:2007}, but in order to reduce the uncertainty due the unknown elementary pion electroproduction off a neutron and uncertainties in the Fermi smearing corrections, the pion nuclear transparency was later redefined~\cite{Qian:2010} as the ratio of pion electroproduction cross sections from the nuclear target to those from the deuteron. The deuterium nuclear transparency was found to be independent of $P_{\pi}$ (or $Q^2$) with 81\% probability, hence, both methods yielded almost identical nuclear transparencies. The extracted transparency as a function of the pion momentum $P_{\pi}$ for all targets is shown in Fig.~\ref{fig:pict} (left panel).

The measured pion nuclear transparencies are compared to three different calculations. Although, all three 
calculations use an effective interaction based on the quantum diffusion model~\cite{FLFS88} to incorporate the 
CT effect, the underlying conventional nuclear physics is calculated very differently. The calculations of  
Larson, {\it et  al.}~\cite{Larson:2006ge}, use a semi-classical formula based on the Eikonal approximation, 
Cosyn {\it et al.} use a relativistic multiple-scattering Glauber approximation (RMSGA) integrated over the 
kinematic range of the experiment and compare it to a relativistic plane wave impulse approximation (RPWIA) to 
calculate the nuclear transparency. Finally, Kaskulov, {\it et al.}~\cite{mosel2} use a model built 
around a microscopic description~\cite{Mosel} of the elementary $^1$H($e,e'\pi^+$)n process, which is divided 
into a soft hadronic part and a hard partonic or a deep inelastic scattering production part. For the reaction on 
nuclei, the elementary interaction is kept the same and nuclear effects such as Fermi motion, Pauli blocking and 
nuclear shadowing, are incorporated. Finally, all produced pre-hadrons and hadrons are propagated through the 
nuclear medium according to the Boltzmann-Uehling-Uhlenbeck (BUU) transport equation. The nuclear transparency is calculated as the ratio of the differential cross section calculated in this model, with and without FSI. 

The observed pion nuclear transparency (as compared both to hydrogen and deuterium cross sections) shows a steady rise versus pion momentum for the nuclear targets, causing a deviation from calculations which do not include CT. Also, measured rises in nuclear transparency versus $Q^2$ are consistent with the rise in transparency in all three calculations that include CT, even though the underlying cause for the rise in nuclear transparency is different for the different model calculations.

The $A$ dependence of the nuclear transparency gives further insight on the proper interpretation of the data in terms of an onset of CT. The entire nuclear transparency data set was examined using a single parameter fit to $T=A{\alpha (Q^2) -1}$, where $A$ is the nuclear mass  number and $\alpha (Q^2)$  is the free parameter. Even though this single-parameter fit is simplistic and neglects local A-dependent shell or density effects, it does not affect the final conclusion that the A-dependence changes with $Q^2$. Thus, even though the exact value of $\alpha$ may come with a variety of nuclear physics uncertainties, a significant empirical $Q^2$ dependence is observed from the 
data. In Fig. \ref{fig:pict} (right panel), we compare $\alpha$ as function of $Q^2$, extracted from the single 
parameter form $T=A^{\alpha(Q^2) -1}$,  along with the calculations including CT effects of Larson, {\it et al.}~\cite{Larson:2006ge} and Cosyn, {\it et al.}~\cite{Cosyn:2006}.

The results of the pion electroproduction experiment demonstrate that both the energy and $A$ dependence of the 
nuclear transparency show a significant deviation from the expectations of conventional nuclear physics and are 
consistent with calculations that include CT. The results can be seen as a clear indication of the onset of 
CT for pions.

\subsubsection{Rho Electroproduction}
\label{sec:rho}
Electroproduction of vector mesons from nuclei is another excellent tool to investigate the 
formation and propagation of quark-antiquark ($q\bar{q}$) pairs under well-controlled 
kinematical conditions. These $q\bar{q}$ states of mass M$_{q\bar{q}}$ can propagate over a 
distance $l_c$ known as the coherence length and given by $l_c=2\nu/(Q^{2}+M_{q\bar{q}}^2)$, 
where $-Q^2$ and $\nu$ are the squared mass and energy of the photon in the lab frame 
(for reviews and references see e.g \cite{Bauer:1977iq,Piller:1999wx}). The HERMES collaboration 
at DESY \cite{Ackerstaff:1998wt} used exclusive incoherent electroproduction off 
$^1$H and $^{14}$N to study the interaction of the $q\bar{q}$ fluctuation 
with the nuclear medium by measuring the nuclear transparencies of  $^{14}$N relative to $^1$H 
as a function of the coherence length $l_c$. They found a coherence length dependence of 
the nuclear transparency of $^{14}$N that is consistent with the onset of hadronic 
initial state interactions where the $q\bar{q}$ pair interacts with the nuclear medium 
like a $\rho^0$ meson. When the coherence length $l_c$ is smaller than the mean free path of 
the $\rho^0$ meson in the nuclear medium, it is expected that the initial state interaction 
of the $q\bar{q}$ is predominantly electromagnetic and thus the nuclear transparency is 
independent of $l_c$. The probability of the $q\bar{q}$ pair to interact with the nuclear medium 
increases with $l_c$ until $l_c$ exceeds the nuclear size \cite{Hufner:1996dr}. The HERMES 
measurements have important implications for the study of color transparency using $\rho^0$ 
meson electroproduction, where the CT signal would be the increase of the nuclear transparency 
with Q$^2$, which controls the initial size of the $\rho^0$ meson. These results demonstrate 
that the increase of the nuclear transparency when $l_c$ decreases (Q$^2$ increases) can mimic 
the CT effects. Therefore, to unambiguously identify the CT signal, one should keep 
$l_c$ fixed while measuring the Q$^2$ dependence of the nuclear transparency, or perform the 
measurements in the regions where no $l_c$ dependence is expected.

When CT effects are present, a photon of high virtuality $Q^2$ is expected to produce a 
$q\bar{q}$ pair with small $\sim 1/Q^2$ transverse separation, which will have reduced 
interaction in the nuclear medium. The dynamical evolution of this small size 
colorless $q\bar{q}$ pair to a normal size $\rho^0$ is controlled by the time or length scale called 
formation time $t_f$ or formation length $l_f = c~t_f$ given by $l_f = 2\nu/({m_{v'}}^2 - {m_{v}}^2)$, 
where $m_v$ is the mass of the $\rho^0$ in the ground state and $m_{v'}$ is the mass of its 
first radial excitation. The first measurements to study CT effects using incoherent 
diffractive $\rho^0$ leptoproduction off nuclei were performed at FNAL by the E665 
collaboration \cite{Adams:1995} and CERN by the NMC collaboration \cite{Arneodo:1994id}. 
Both experiments used muon beams with 450 GeV and 200 GeV energy, respectively. At these high 
energies, both $l_f$ and $l_c$ become larger than the nuclear radius and therefore coherence 
length effects are not expected to play any role in the CT signal because the fluctuations of 
the transverse size of the $q\bar{q}$ pair are ``frozen'' during the propagation. Although the 
two measurements are consistent with each other, the NMC experiment reaches Q$^2$ values close to 10 (GeV/c)$^2$, and shows no Q$^2$ dependence, while the E665 experiment measures an increase of the nuclear transparency with Q$^2$. In addition, E665 
performed $Q^2$-dependent fits to the exclusive $\rho^0$ production cross section, $\sigma_A$, 
from hydrogen, deuterium, carbon, calcium and lead by a power law $\sigma_A = \sigma_0 A^\alpha$ 
in which $\sigma_0$ and $\alpha$ are parameters. At low $Q^2$, the value of $\alpha$ measured is 
compatible with 2/3, a value characteristic of soft nuclear interactions but an increase in $\alpha$ was observed at
higher $Q^2$. Due to the limited statistical precision of these results they are only  suggestive of CT effects.

The HERMES collaboration  measured nuclear transparency as a function of the coherence length $l_c$ for incoherent electroproduction of $\rho^0$ mesons off a nucleon in a nucleus as well as for coherent $\rho^0$ production on a nucleus as a whole~\cite{Airape:2003}. The nuclear transparency for coherent production was found to increase with the coherence length, as expected from the effects of the nuclear form factor 
\cite{Kopeliovich:2001xj}. In order to study CT effects, the $Q^2$ dependence of the nuclear transparency was measured at fixed $l_c$~\cite{Airape:2003}. The nuclear transparencies were extracted for both coherent and incoherent production in each bins of ($l_c$, Q$^2$). Due to the low statistics, the data were fitted with a common $Q^2$-slope $(P_1)$, which was extracted assuming $T_{(coh/inc)} = \sigma^{^{14}N}_{(coh/inc)}(l_c,Q^{2}) / A \sigma_{p} = P_0 + P_1\cdot Q^{2}$, letting $P_0$ vary independently in each $l_c$ bin and keeping $P_1$ as common free parameter. The common slope parameter of the Q$^2$-dependence, $P_1$, has been considered as  the signature of the CT effect averaged over the coherence length range. The Q$^2$ slopes for coherent and incoherent productions were found to be ($0.070 \pm 0.021$) and ($0.089 \pm 0.046$) (GeV/c)$^{-2}$, respectively in agreement with model calculations \cite{Kopeliovich:2001xj}. Although these slopes were found to be positive and 
as such consistent with CT effects, the statistical significance of these results was limited.

\begin{figure}[th]
\centerline{\psfig{file=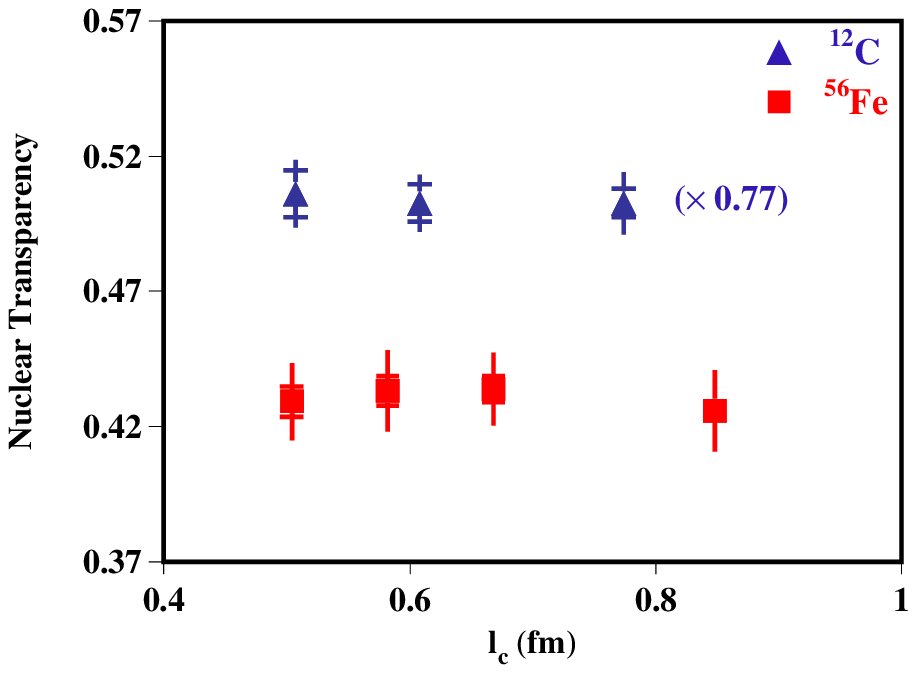,width=7cm}\psfig{file=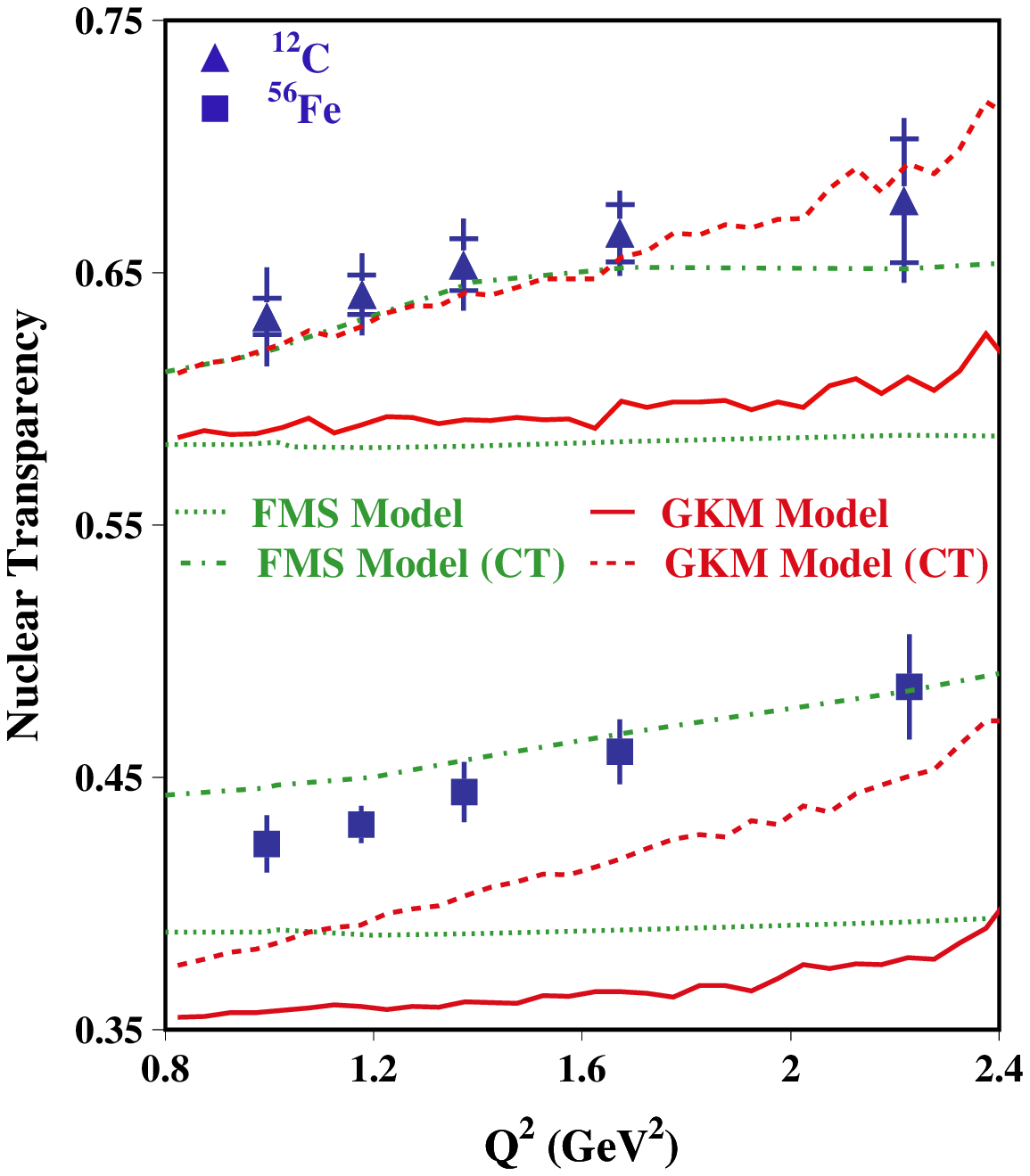,width=7.5cm}}
\vspace*{8pt}
\caption {(Left panel) Nuclear transparency as a function of $l_c$. The carbon 
data has been scaled by a factor 0.77 to fit in the same figure with the iron data \protect\cite{ElFassi:2012nr}. (Right panel) Nuclear transparency as a function of $Q^2$. The curves are predictions of the FMS \protect\cite{Frankfurt:2008pz} (red) and GKM \protect\cite{Gallmeister:2010wn} (green) models with (dashed-dotted and dashed curves), respectively) and without (dotted and solid curves, respectively) CT. Both models include the pion absorption effect when the $\rho^0$ meson decays inside the nucleus. The inner error bars are the statistical uncertainties and the outer ones are the statistical and the point-to-point systematic uncertainties added in quadrature.}
\label{fig:clas-ct}
\end{figure}

Recently, CLAS collaboration measured the nuclear transparency for incoherent exclusive 
$\rho^0$ electroproduction off carbon and iron relative to deuterium \cite{ElFassi:2012nr} 
using a 5 GeV electron beam. Both the deuterium target and the solid target (carbon, iron) 
were exposed to the beam simultaneously to reduce systematic uncertainties in the nuclear 
ratio and allow high precision measurements. The $\rho^0$ mesons were identified through the 
reconstructed invariant mass of the two detected pions with 0.6 $< M_{\pi^+ \pi^-} < $ 1 GeV. A set of kinematic 
conditions were imposed to identify exclusive diffractive and incoherent $\rho^0$ events, and the $t$ 
distributions for exclusive events were fit with an exponential form $Ae^{-bt}$. The slope 
parameters $b$ for $^2$H (3.59 $\pm$ 0.5), C (3.67 $\pm$ 0.8) and Fe (3.72 $\pm$ 0.6) were reasonably 
consistent with the hydrogen measurements~\cite{Morrow:2008ek}  of 2.63 $\pm$ 0.44 taken with 
5.75 GeV beam energy. The transparencies for C and Fe are shown as a function of $l_c$ in Fig.~\ref{fig:clas-ct}. 
As expected, they do not exhibit any $l_c$ dependence because $l_c$ is much shorter than the C and Fe nuclear 
radii of 2.7 and 4.6 fm respectively. Consequently, the coherence length effect cannot mimic 
the CT signal in this experiment. Fig.~\ref{fig:clas-ct} shows the increase of the transparency with $Q^2$ for both C 
and Fe, indicating the onset of CT phenomenon. The rise in transparency with $Q^2$ corresponds to 
an $(11 \pm 2.3)\%$ and $(12.5 \pm 4.1)\%$ decrease in the absorption of the $\rho^0$ in Fe and C respectively. 
The Q$^2$ dependence of the transparency was fitted by a linear form $T_A = a~Q^2 + b$.

The extracted slopes ``$a$'' for C and Fe are in good agreement with both Kopeliovich-Nemchik-Schmidt (KNS) \cite{Kopeliovich:2007mm} and Gallmeister-Kaskulov-Mosel (GKM~ \cite{Gallmeister:2010wn} predictions, but somewhat larger than the Frankfurt-Miller-Strikman (FMS) \cite{Frankfurt:2008pz} calculations. While the KNS and 
GKM models yield an approximately linear $Q^2$ dependence, the FMS calculation yields a more complicated $Q^2$ dependence as shown in Fig.~\ref{fig:clas-ct} (right). The measured slope for carbon corresponds to a drop in the absorption of the $\rho^0$ from 37\% at $Q^2$ = 1 GeV$^2$ to 32\% at $Q^2$ = 2.2 GeV$^2$, in reasonable agreement with the calculations. The measured slopes both in CLAS and HERMES are fairly well described  by the KNS model. 
The FMS model is quite successful in reproducing both the slopes and the magnitudes of the nuclear transparencies, while taking into account both CT effect and the $\rho^0$ decaying inside the nucleus and the subsequent pion absorption effect. The same model is successful in reproducing the JLab pion electroproduction data discussed in section \ref{sec:pion}. 

The onset of CT in $\rho^0$ electroproduction seems to occur at lower Q$^2$ than in the 
pion measurements. This early onset suggests that diffractive meson production might be the optimal way to 
create small size $q\bar{q}$ pair. The $Q^2$ dependence of the transparency ratio is mainly sensitive 
to the reduced interaction of the $q\bar{q}$ pair as it evolves into a full-sized hadron, and 
thus depends strongly on the formation time during which the small size configuration's 
color fields expand to form a $\rho^0$ meson. The formation time used by the FMS and GKM models 
is between 1.1 and 2.4 fm for $\rho^0$ mesons produced with momenta from 2 to 4.3 GeV while the 
KNS model uses an expansion length roughly a factor of two smaller. The agreement between 
the observed $Q^2$ dependence and these models suggests that these assumed expansion 
distances are reasonable. Having established these features, detailed studies of the 
theoretical models will allow the first quantitative evaluation of the structure and evolution 
properties of the small size configurations. Such studies will be further enhanced by future measurements \cite{hafidi06}, which will include additional nuclei and extend to higher $Q^2$ values.

\section{Future Experiments}
There are already approved  plans for extending CT studies of the $A(e,e'p)$ and $A(e,e'\pi)$ reactions to much higher energies following the upgrade of JLab to 12 GeV. This will finally allow to reach kinematics where $l_c$ is larger than the interaction length for a nucleon/pion in the nuclear media. The extension of the $A(e,e'p)$ experiment will double the $Q^2$ range covered from the current $Q^2$ = 8.0 (GeV/c)$^2$ to $Q^2$ = 16.0 (GeV/c)$^2$. 
At these higher $Q^2$ values, CT predictions diverge appreciably from the predictions of conventional calculations. As mentioned earlier the BNL $A(p,2p)$ data seem to establish a definite increase in nuclear transparency for nucleon momenta between about 6 and 10 GeV/c. For $A(e,e'p)$ measurements comparable momenta of the ejected nucleon correspond to about 10 $< Q^2 <$ 17 (GeV/c)$^2$, exactly the range of the proposed extension. Hence, this would unambiguously answer the question whether one has entered the CT region for nucleons, and help establish
the threshold for the onset of CT phenomenon in three-quark hadrons. 

The extension of the $A(e,e'\pi)$ experiment will also double the $Q^2$ range covered from the current $Q^2$ = 5.0 (GeV/c)$^2$ to $Q^2$ = 10.0 (GeV/c)$^2$. A $Q^2$ dependence of the pion transparency in nuclei may also be 
introduced by conventional nuclear physics effects at the lower $Q^2$. Thus one must simultaneously examine both the $Q^2$ and the $A$ dependence of the meson transparency. Several independent calculations~\cite{ralston,Larson:2006ge} predict the CT effect to be largest around $Q^2$ of 10 (GeV/c)$^2$, which is in agreement with the observation of full CT in the Fermilab experiment mentioned above. Using the data collected at 6 GeV as a baseline, the new data could help confirm and establish the CT phenomenon in mesons on a firm footing. 

The JLab 12 GeV $A(e,e'\rho^0)$ experiment \cite{hafidi06} will extend the maximum $Q^2$ reach from 2.2 to 5.5 (GeV/c)$^2$, which will allow for significant increase in the momentum and energy transfer involved in the reaction. Therefore, one expects to produce smaller configurations that live longer: the optimum parameters for CT studies. Several nuclei including deuterium, carbon, iron and tin will be studied. Measurements with different nuclei sizes are important for quantitative understanding of the small size configuration's formation time and its interaction in the nuclear medium. The dependence of the nuclear transparency on the coherence length will be measured for $l_c$ range up to 2.5 fm. The measurements will be performed for fixed coherence length. 

A complementary strategy is to use processes where multiple rescattering dominates in light  nuclei ($^2$H, $^3$He) which allows to suppress the expansion effects. An additional advantage of these processes is that one can use for the calculations generalized Eikonal approximation, see review in \cite{Sargsian:2001ax}. In particular,  these reactions are well suited to search for a precursor  of CT - suppression of the configurations in nucleons with  pion cloud in the hard processes like the nucleon form factors at relatively small $Q^2 \ge $ 1 (GeV/c)$^2$ - chiral transparency~\cite{Frankfurt:1996ai}. The simplest reaction of this kind is production of a slow $\Delta$ isobar in the
 process $e + D \to e + p + \Delta^0$ which should be suppressed in the chiral transparency regime.

\section{Conclusions}

Color Transparency is a key property of QCD. It offers a unique probe of ``color'', a defining feature of QCD, yet totally invisible in the observed structure of ordinary nuclear matter. CT is well established at very high energies, where the small size configuration is highly relativistic and its lifetime in the nucleus rest frame is dilated, causing it to stay small while traversing the nucleus. At low and intermediate energies, the situation is more challenging because the small size configuration starts expanding inside the nucleus. However, studying CT at low and intermediate energies provides valuable information on the small size configuration formation, expansion dynamics and most importantly, its interactions with the nuclear medium as a function of its color field. Furthermore, the onset of CT is a necessary condition for factorization, which is an important requirement for accessing GPDs in deep exclusive meson production.
Important experimental efforts have been dedicated to the search for CT both at high and low/medium energies. No evidence for CT in the baryon sector was observed while measurements in the meson sector combined can definitely be considered as a strong evidence for the CT phenomenon. One should point out the latest pion and rho meson measurements from Jefferson lab. Establishing the onset of CT phenomenon is just the beginning. The next step, which will be done at Jefferson Lab 12 GeV, is to understand quantitatively the small size configuration formation time and its interaction in the nuclear environment. Therefore, one needs to extend the $Q^2$ range, which will allow for a significant increase in the momentum and energy transfer involved in the reaction. The measurements on several nuclei with different sizes will allow studying the space-time properties of these small size configurations during their evolution to full size hadrons.

\section*{Acknowledgments}
This work was supported by the U.S. Department of Energy, Office of Nuclear Physics, under contracts No.~DE-AC02-06CH11357 and ~DE-FG02-07ER41528.

\end{document}